# Efros-Shklovskii variable range hopping in reduced graphene oxide sheets of varying carbon $sp^2$ fraction

Daeha Joung[1,2] and Saiful I. Khondaker[1,2,3] *

[1] Nanoscience Technology Center, University of Central Florida, Orlando, Florida 32826, USA
[2] Department of Physics, University of Central Florida, Orlando, Florida 32826, USA
[3] School of Electrical Engineering and Computer Science, University of Central Florida, Orlando, Florida 32826, USA

* To whom correspondence should be addressed. E-mail: saiful@ucf.edu

We investigate the low temperature electron transport properties of chemically reduced graphene oxide (RGO) sheets with different carbon $sp^2$ fractions of 55 to 80 %. We show that in the low bias (Ohmic) regime, the temperature ($T$) dependent resistance ($R$) of all the devices follow Efros-Shklovskii variable range hopping (ES-VRH) $R \sim \exp[(T_{ES}/T)^{1/2}]$ with $T_{ES}$ decreasing from 30976 to 4225 K and electron localization length increasing from 0.46 to 3.21 nm with increasing $sp^2$ fraction. From our data, we predict that for the temperature range used in our study, Mott-VRH may not be observed even at 100 % $sp^2$ fraction samples due to residual topological defects and structural disorders. From the localization length, we calculate a bandgap variation of our RGO from 1.43 to 0.21 eV with increasing $sp^2$ fraction from 55 to 80 % which agrees remarkably well with theoretical prediction. We also show that, in the high bias regime, the hopping is field driven and the data follow $R \sim \exp[(E_0/E)^{1/2}]$ providing further evidence of ES-VRH.

## I. INTRODUCTION

Chemical functionalization of graphene has attracted significant research interests due to its potential in obtaining a bandgap in graphene and thereby tuning the electrical properties from semimetal to insulator.[1-17] In particular, solution processed route for producing reduced graphene oxide (RGO) sheets, which has a wide range of oxygen functionalities such as hydroxyl and epoxy groups, received great attention due to its (i) high throughput manufacturing, (ii) tunable electrical and optical properties via controlling the ratio of $sp^2$ C-C and $sp^3$ hybridized carbon (i.e., oxygen functional groups) and (iii) ability to anchor different types of nanoparticles and organic molecules, which pave the way for potential applications in flexible electronics, photovoltaics, supercapacitors and battery.[1, 2, 15, 18-28]

Functionalization of graphene creates disorders and the low temperature electronic transport properties of these structures are akin to that of disordered semiconductors where electron localization and hopping conduction play a significant role. However, a clear understanding of the electronic transport properties of the RGO sheets is lacking as different study reports different conduction mechanisms such as Mott variable range hopping (VRH) and Efros-Shklovskii (ES-) VRH.[1, 10, 29-32] Understanding of the electron transport properties of RGO is of great significance to realize the overreaching goals of functionalized graphene and its composites. The difference between the Mott and ES-VRH is in the details of their localization parameters, density of states (DOS) and interactions that manifest in the temperature dependence of resistance (R).[33-40] In general, the VRH can be characterized as

$$R(T) = R_0 \exp\left(\frac{T_0}{T}\right)^p \qquad (1)$$

where $R_0$ is a prefactor, $T_0$ is a characteristic temperature and $p$ is a characteristic exponent the value of which distinguishes different conduction mechanism. Since the hopping conduction occurs between the localized states around the Fermi level ($E_F$), the details of the DOS around $E_F$ is an important consideration in determining the temperature dependence of resistance. Mott considered a constant DOS and showed that the value of $p$ in Eq. (1) is given by $p = 1/(D+1)$, where $D$ is the dimensionality of the system under investigation.[33, 34] Therefore in Mott-VRH, $p = 1/3$ for 2D system. The characteristic temperature for Mott-VRH in 2D is then given by

$$T_0 \equiv T_M = \left(\frac{3}{k_B N(E_F)\xi^2}\right) \qquad (2)$$

where $N(E_F)$ is the DOS near $E_F$ and $\xi$ is the localization length. However, Efros and Shklovskii later pointed out that, at low enough temperature, the DOS near the $E_F$ is not constant rather it vanishes linearly with energy for a 2D system.[35, 36, 38] This is because, when an electron hops from one site to another, it leaves a hole and the system must have enough energy to overcome this electron-hole Coulomb interaction. This vanishing DOS, called Coulomb gap ($E_{CG}$), results in the temperature dependence of resistance, that can still be described with Eq. (1) but with $p = 1/2$ in all dimension. The characteristic temperature in 2D then becomes

$$T_0 \equiv T_{ES} = \left(\frac{2.8e^2}{4\pi\varepsilon\varepsilon_0 k_B \xi}\right) \qquad (3)$$

where $\varepsilon_0$ and $\varepsilon$ are the value for permittivity of vacuum and the dielectric constant of the material. For some samples, the disorder may be very high so that $E_{CG}$ is dominant at all measureable temperatures giving ES-VRH only. On the other hand, in other relatively low disordered samples, the energy scale is such that the carriers may have enough energy to overcome $E_{CG}$ at all measurable temperatures, which means the DOS is practically constant. In that case, only Mott-VRH will be dominant.[35] At intermediate disorders, it may be possible to see a crossover from ES to Mott-VRH with increasing temperature in the same sample.

Additional evidence of ES-VRH can also be obtained from electric field dependent transport study at a fixed temperature. Since the energy necessary for hopping can also be obtained from the electric field ($E$) rather than temperature, at high enough electric field (high bias regime) the temperature dependence is strongly reduced and one enters the regime of field driven hopping transport, where the conduction is given by [39-43]

$$R(E) \sim \exp\left(\frac{E_0}{E}\right)^{1/2}, \qquad (4)$$

$$\text{with } E_0 = \frac{2k_B T_{ES}}{e\xi}. \qquad (5)$$

where $T_{ES}$ and $\xi$ represent the same parameters as in Ohmic ES-VRH of Eq. (3).

In this paper, we present detailed temperature (295 to 4.2 K) and field dependent electron transport investigations of RGO sheets with different degrees of carbon $sp^2$ fraction. The carbon $sp^2$ fraction was tuned from 55 to 80 % by varying reduction time in hydrazine hydrate reduction method. The devices with channel length and width of 500 nm × 500 nm were fabricated by dielectrophoretic (DEP) assembly of RGO sheets. In the low bias Ohmic regime, we show that the temperature dependence of resistance follows ES-VRH model $R = R_0 \exp[(T_{ES}/T)^{1/2}]$ for all RGO devices with $T_{ES}$ decreasing from 30976 to 4225 K and $\xi$ increasing from 0.46 to 3.21 nm with increasing carbon $sp^2$ fraction. Interpolating the data to 100% carbon $sp^2$ fraction, we predict that for the temperature range used in our study, Mott-VRH may not be observed even at 100 % carbon $sp^2$ fraction possibly because of residual topological defects and structural disorders. From the localization length, we calculate a bandgap variation of our RGO from 1.43 to 0.21 eV with increasing $sp^2$ fraction from 55 to 80 % which agrees remarkably well with theoretical prediction. At low temperature and high electric field (high bias regime), our data can be explained with field dependent ES-VRH model $R \sim \exp[(E_0/E)^{1/2}]$, providing further evidence of ES-VRH in our samples. With increasing carbon $sp^2$ fraction, the measured values of $E_0$ decreased from $16.1 \times 10^8$ to $1.38 \times 10^8$ V/m. These values are in qualitatively agreement with calculated $E_0$ from Ohmic ES-VRH.

## II. EXPERIMENTAL DETAILS
### A. Synthesis of RGO sheets with different carbon s$p^2$ fraction (reduction efficiency)

RGO sheets used in this study were obtained via chemical reduction of individual graphene oxide (GO) sheets. The individual GO sheets in powder form were obtained from Cheaptubes InC.[44] 15 mg of GO powder was added to a flask containing 15 mL of deionized (DI) water. Then, the GO solution was stirred with a Teflon-coated magnetic stirring bar in a water bath for 24 hours to obtain a good dispersion. The average lateral dimension of the GO sheets was about ~ 0.8 μm and the average thickness was ~ 1 nm indicating single layer GO sheet.[20] 100 μL of 5 % ammonia aqueous solution and 15 μL of hydrazine hydrate (Sigma-Aldrich St. Louis, MO, 35 % DMF) were added to the GO solution. The mixture was then heated at 90 °C for either 10, 20, 30, 45 or 60 minutes under stirring to produce RGO sheets of different reduction efficiency. Another mixture was left in hydrazine for 24 hours without any heating. The reduction efficiency was determined from carbon $sp^2$ fraction using X-ray Photoelectron Spectroscopy (XPS).

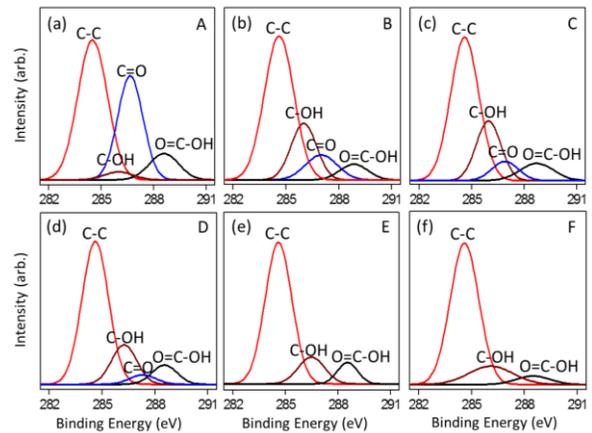

Figures 1(a)-(f) show deconvolution of the *C1*'s peak in the XPS spectrum of the RGO sheets of different reduction efficiency. Figure 1(a) (sample A) represents the resulting RGO sheet that was not heated, while Figs. (b), (c), (d), (e) and (f) represent the resulting RGO sheets (defined as B, C, D, E and F) obtained from the different heating (reduction) time for either 10, 20, 30, 45 or 60 min, respectively. The four deconvoluted peaks indicate the deoxygenated graphene C-C at 284.6 ± 0.1 eV, oxygen-containing functional groups for hydroxyl (C-OH) at 286.0 ±0.1 eV, carbonyl (C=O) at 287.0 ±0.2 eV, and carboxyl acid (O=C-OH) at 288.6 ± 0.1 eV.[45-47] The C-C peak refers to the amount

FIG. 1. (Color online) Deconvolution of the C1's peaks of XPS spectra for different reduction efficiency of RGO sheets. The reduction time was (a) 0, (b) 10, (c) 20, (d) 30, (e) 45 and (f) 60 min. The peaks containing different groups C-C, C-OH, C=O and O=C-OH are labeled for clarity.

of $sp^2$ carbon components, while the oxygen-containing functional groups located on the basal plane of the sheets and the edges of the sheets refer to the amount of $sp^3$-hybridized carbon.[6, 29, 48, 49] Since the presence of $sp^3$ defect sites distorts the intrinsic $\pi$ state of the $sp^2$ sites,[1, 15, 49-51] residual carbon $sp^2$ fraction is an important clue for RGO sheets and regarded as a reduction efficiency. The carbon $sp^2$ fraction was calculated by taking the ratio of the integrated peak areas corresponding to the C-C peak to the total area under the $C1$'s spectrum. The percentage of the carbon $sp^2$ fraction can be determined by the following expression:

$$\frac{A_{C-C}}{A_{C-C} + A_{C-OH} + A_{C=O} + A_{O=C-OH}} \times 100\% \quad (6)$$

where A denotes the area under the corresponding peaks as maked in Figs. 1(a)-(f). The carbon $sp^2$ fractions are 55, 61, 63, 66, 70 and 80 % for A, B, C, D, E, and F, respectively. This result indicates that the carbon $sp^2$ fraction (or reduction efficiency) of RGO sheets increases with increasing reduction time.

## B. Device fabrication and measurement set up

Devices were fabricated on heavily doped silicon (Si) substrates capped with a thermally grown 250 nm thick $SiO_2$ layer. Source and drain electrode patterns of 500 nm × 500 nm (channel length × width) were defined by electron beam lithography (EBL) followed by thermal deposition of 3 nm thick Cr and 25 nm thick Au. The RGO sheets were then assembled between the prefabricated source and drain electrodes using *AC* dielectrophoresis (DEP). Details of the DEP device assembly can be found in our previous publication.[20] In brief, a 3 μL of RGO solution was drop casted onto the electrode pattern. An *AC* voltage of 3 $V_{p-p}$ with a frequency of 1 MHz was applied between the source and drain electrodes for 1 minute. After the DEP assembly, atomic force microscope (AFM) was used to characterize the RGO devices. Figure 2(a) shows a tapping-mode AFM image of a representative device along with its height analysis. From this figure, it can be seen that the thickness varies from 2 to 7 nm in the channel, indicating that up to seven layers of RGO sheets have been assembled. The thickness of RGO sheets in the channel is varied between 5 and 15 nm. This is typical for all of our devices.

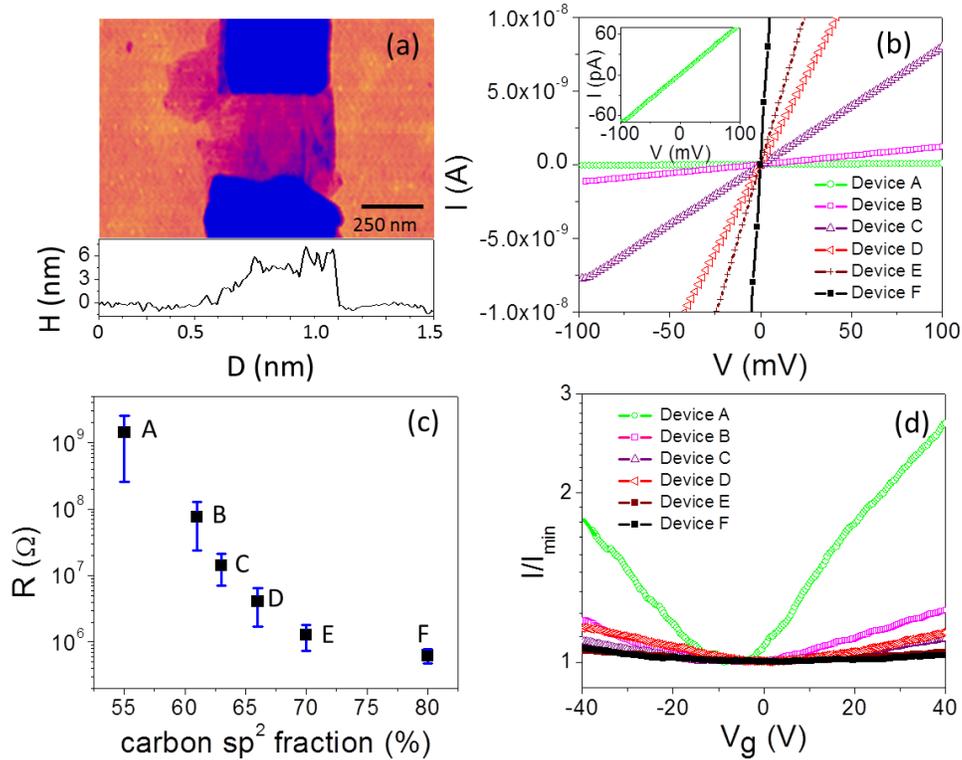

FIG. 2. (Color online) (a) Tapping mode atomic force microscope (AFM) image of a RGO device along with its height profile. (b) Room temperature current – voltage ($I$-$V$) characteristics of RGO devices with different carbon $sp^2$ fraction. Inset shows zoomed in $I$-$V$ for device A. (c) Room temperature resistance ($R$) of RGO sheets with different carbon $sp^2$ fraction. (d) current-gate voltage ($I$-$V_g$) characteristics of all RGO devices with fixed bias voltage of 1V. For clarity, the current was normalized to its minimum current $I_{min}$.

The devices were then bonded to a chip carrier and loaded into a variable temperature cryostat for temperature-dependent electronic transport measurements. The measurements were performed using a Keithley 2400 source meter, and a current preamplifier (DL 1211) capable of measuring pA signal interfaced with the LABVIEW program. For each carbon $sp^2$ fraction, we have measured ~ 20 devices.

### III. RESULTS AND DISSCUSSION

Figure 2(b) shows the representative room temperature current-voltage (*I-V*) characteristics of RGO devices A, B, C, D, E and F containing different carbon $sp^2$ fraction. Within the voltage range of -100 to 100 mV, the *I-V* curves are Ohmic allowing us to calculate the resistance of the samples. For each $sp^2$ fraction, we measured resistance values of 20 samples. The average room temperature resistance (*R*) of the devices is presented in Fig. 2(c) with their corresponding carbon $sp^2$ fraction. The decrease in carbon $sp^2$ fraction resulted in increase of *R* (or decrease conductivity). *R* for device A is ~ $1.06 \times 10^9$ Ω while for device F it is ~ $0.6 \times 10^6$ Ω demonstrating that the value of *R* can be tuned by more than 3 orders of magnitude but tuning the carbon $sp^2$ fraction from 55 to 80%. The decrease of resistance with increasing $sp^2$ fraction demonstrates that restoration of π-π bond improves charge percolation pathways in the RGO sheet. However, we note that initially the decrease of resistance with increasing $sp^2$ fraction (55 to 70%) is more dramatic and then it started to level off above 70 %. This is due to the fact that, even though the π-π bonds are restored, transmission electron microscopy (TEM) images from Erickson et al. and Gómez-Navarro et al. shows that such improvement occurs at the expense of increasing topological defects.[52, 53] So we believe, at about 70 % $sp^2$ fraction, topological defects started to play a major role in resistance than the remaining $sp^3$ fraction. In other words, even if we are able to reduce the sample such that $sp^2$ fraction is close to 100 %, the *R* of RGO will not come close to graphene due to the residual topological defects. In Fig. 2(d) we present representative room temperature current – back-gate voltage (*I - $V_g$*) curves for sample A to F measured from -40 to + 40 V at a fixed bias voltage of 1 V. For clarity, the current was normalized to its minimum current $I_{min}$. Typical ambipolar characteristics are observed for all devices with highest current on-off occurring for lowest $sp^2$ fraction, as expected.[10]

In order to determine the hopping conduction mechanisms, we measured temperature dependence of *R*. Figure 3(a) shows semi-log scale plot of *R* versus (vs) *T* for samples A, B, C, D, E and F containing different carbon $sp^2$ fraction. The values of *R* for each sample was measured at a fixed low bias voltage of ~ 100 mV when the temperature was lowered from 295 to 4.2 K at a rate of 0.04 K/s. We observed non-Ohmic behavior below 200 K for device A and below 40 K for device F within the voltage range of 100 to + 100 mV. This is more clearly seen in Figs. 3(b) and (c) where we show the *I-V* characteristic at a few selected temperatures measured from -100 to + 100 mV for device A and device F respectively. Since *R* is defined from the Ohmic part of the *I-V* curve, in Fig. 3(a) we discarded data below those temperatures that did not have linear *I-V* curves at 100 mV. In addition, the resistance measured from the *I-V* curve at a few selected temperatures agrees well with the resistance values plotted in Fig. 3(a) indicating the accuracy of the data. We also note that except for device A, the resistance for all of our samples varied from 2 to more than 3 orders of magnitude with temperature. Such a large variation is important for accurate analysis of hopping conduction.

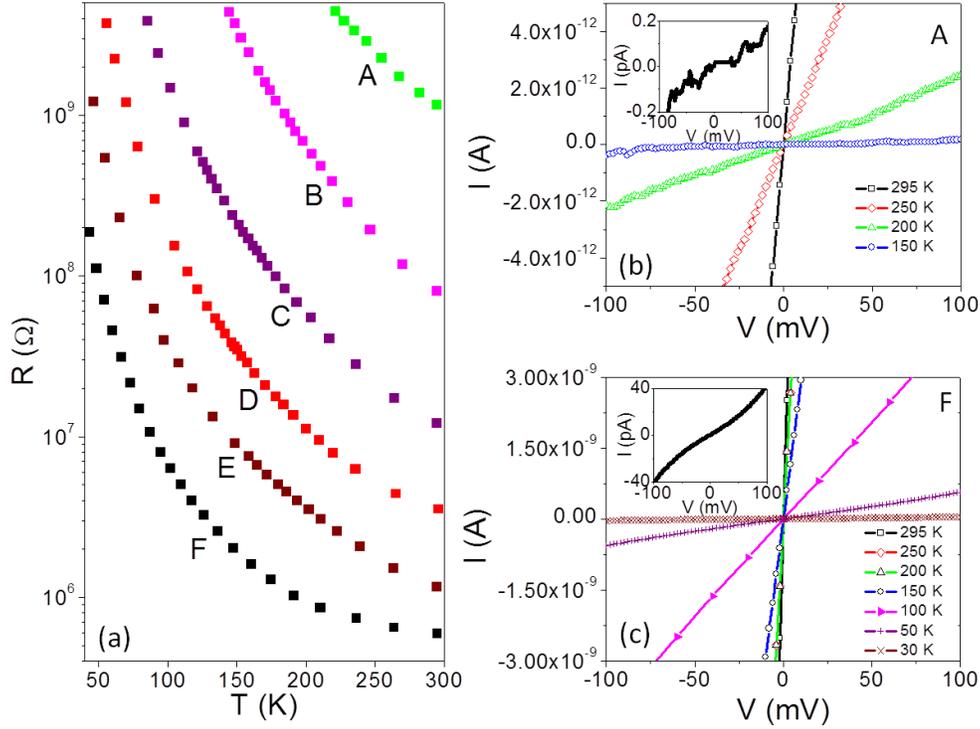

FIG. 3. (Color online) (a) Semi-log scale plot of resistance (*R*) versus (vs) temperature (*T*) for samples A, B, C, D, E and F in the temperature range of 295-40 K. (b) *I-V* characteristics of device A in the temperature range of 295-150 K at bias voltage range from -100 to + 100 mV. Inset shows *I-V* at 150 K. (c) *I-V* characteristics of device F in the temperature range of 295-30 K. Inset shows zoomed in *I-V* at 30 K.

The usual practice of determining 2D hopping conduction mechanism is by plotting *ln R* vs either $T^{1/3}$ (Mott-VRH) or $T^{1/2}$ (ES-VRH). Most work on RGO only showed a plot of *lnR* vs $T^{1/3}$ claiming Mott-VRH without making any comments whether the data could also be fitted with $T^{1/2}$.[10, 29-31] However, it has been previously reported that often the same data can be fitted with both $T^{1/3}$ and $T^{1/2}$ making it extremely difficult for accurate analysis of hopping conduction.[54, 55] This ambiguity can be avoided by determining the exponent *p* in a self consistent way. From Eq. (1), one can obtain the logarithmic derivative *W*:[37, 38, 55, 56]

$$W = -\frac{\partial \ln \rho(T)}{\partial \ln T} = p \times \left(\frac{T_0}{T}\right)^p. \tag{6}$$

The value of *p* can then be obtained from the slope of *ln W* vs *ln T* plot since *ln W = A – p ln T*. Figures 4(a), (b), (c) and (d) show *ln W* vs *ln T* plot for samples C, D, E and F respectively.

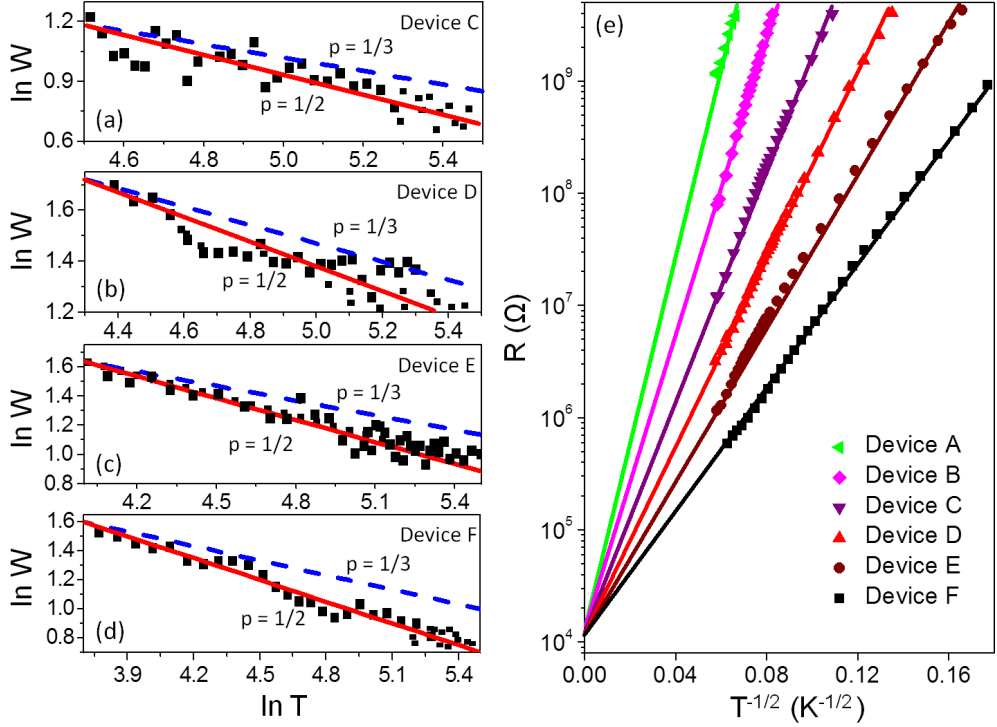

FIG. 4. (Color online) (a)-(d) Reduced activation energy ($W$) plotted vs temperature ($T$) in a log-log scale for device C, D, E, and F, respectively. From the slopes of the plots we obtain $p = 0.464 \pm 0.004$, $0.465 \pm 0.058$, $0.475 \pm 0.001$ and $0.483 \pm 0.004$ for C, D, E, and F corresponding to the ES-VRH for all samples. For a comparison we also show lines with $p = 1/2$ (ES-VRH) and $p = 1/3$ (2D Mott-VRH) for a guide to the eye. (e) Semi-log scale plot of $R$ vs $T^{-1/2}$ for all RGO devices. The symbols are the experimental points and the solid lines are a fit to $T^{-1/2}$ behavior. From the slopes we obtain $T_{ES}$ =30976, 24964, 13280, 8704, 5901 and 4225 K for devices A, B, C, D, E, and F respectively. By extrapolating the solid lines, we determine $R_0$ values of 14.8, 13.6, 14.1, 13.8, 13.1 and 12.6 k$\Omega$ for device A, B C, D, E, and F, respectively.

The symbols are the experimental data points and the solid red lines are a plot of $p = 1/2$ while the dashed lines are a plot of $p = 1/3$ shown for a guide to the eye. It can be clearly seen that for all the samples, the data follow $p = 1/2$ line. In order to determine the accurate values of $p$, we did a least square fit of the data and obtained $p = 0.464 \pm 0.004$, $0.465 \pm 0.058$, $0.475 \pm 0.001$ and $0.483 \pm 0.004$ for C, D, E, and F, respectively. These values are close to 0.5 expected from ES-VRH. We could not do similar analysis for samples A and B due to limited number of data points within a small temperature range. However, since samples A and B are more disordered than C, D, E and F, we can only expect the ES-VRH to dominate there as well. Figure 4(e) shows a semi-log scale plot of $R$ vs $T^{-1/2}$ for all the samples. The symbols are the experimental points and the solid lines are a fit to $T^{-1/2}$ behavior. As expected, the data for all the samples fit very well with $T^{-1/2}$ behavior. By extrapolating solid lines in Fig. 4(e), we obtained the pre-factor $R_0$ values for all RGO sheets. It can be seen that all the traces collapses to almost a single $R_0$ value with a small variation (within experimental error) from 14.8 to 12.6 k$\Omega$. Our self-consistent analysis of finding the value of $p = 1/2$, the excellent fit of $\ln R$ with $T^{-1/2}$ and a nearly universal value of $R_0$ for all RGO samples clearly indicates that there is no conduction mechanism other than the ES-VRH for the entire temperature ranges for all our samples of varying $sp^2$ fraction. This is in clear contrast than the previous report of Mott-VRH in RGO sheets of varying degrees of reduction treatments.[10] The reason could be the limited temperature range used in their study. In addition, the same data might also fit with $T^{-1/2}$. Indeed, we have analyzed some of those results by extracting the points from the graph and found the data also fit very well with $T^{-1/2}$. This suggests that extreme caution should be taken in analyzing temperature dependence data.

From the slopes of Fig. 4(e) we obtain the characteristic temperature $T_{ES}$ for all of our samples. The values of $T_{ES}$ are 30976 K (device A), 24964 K (B), 13280 K (C), 8704 K (D), 5901 K (E) and 4225 K (F) [see solid symbols in Fig. 5 (a)]. From these values of $T_{ES}$ and using Eq. (3), we determine the localization length ($\xi$) to be 0.46, 0.54, 1.03, 1.54, 2.27 and 3.21 nm for samples A, B, C, D, E and F respectively. In determining $\xi$, we used an effective dielectric constant of $\varepsilon = 3.5$ for RGO sheet.[32, 57, 58] Figure 5(b) shows a plot of $\xi$ vs its corresponding carbon $sp^2$ fraction of the sheets. This demonstrates that with increasing $sp^2$ fraction, the localization length increases. This is what is expected. It is well known that RGO consists of ordered graphene domains surrounded by areas of oxidized domains and point defects. It has been estimated from XPS, Raman and TEM studies that the graphitic domain size in RGO can vary from 1 to 6 nm with reduction efficiency.[50, 52, 53, 59, 60] These values are surprisingly closer to our value of $2\xi$ demonstrating that the wave-function is localized inside each graphitic domain. The agreement between localization lengths with the domain size is rather extraordinary given the complexity of the measurements and analysis.

Figures 5(a) shows that, even for our highest reduction sample, $T_{ES}$ is much higher than the room temperature, making it impossible to see Mott-VRH. We extrapolated our data using a second order polynomial fit to see what the $T_{ES}$ will be at 100 % reduction efficiency. We found a value of $T_{ES} = 1800$ K. Similarly we also found a value of $\xi = 7.44$ nm in Fig. 5 (b). These suggest that even at 100 % carbon $sp^2$ fraction, Mott-VRH may not be observed possibly because of residual topological defects and structural disorders.

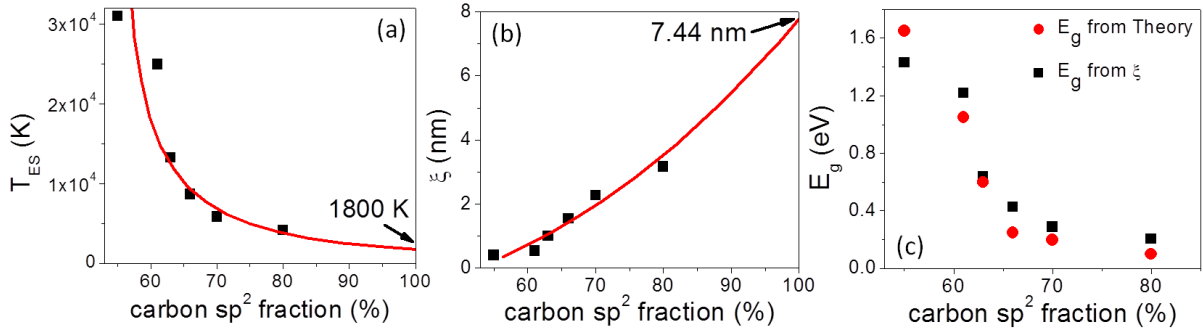

FIG. 5. (Color online) (a) $T_{ES}$ vs their corresponding carbon $sp^2$ fraction of the RGO sheets. The symbols are the experimental points and the red solid lines are extrapolated by a second order polynomial fit. At 100 % $sp^2$ fraction, $T_{ES}$ of 1800 K was detemined. (b) $\xi$ vs their corresponding carbon $sp^2$ fraction of the RGO sheets. At 100 % $sp^2$ fraction, $\xi$ of 7.44 nm was detemined. (c) Bandgap ($E_g$) of RGO samples plotted vs their corresponding carbon $sp^2$ fraction. Square symbols demonstrate $E_g$ calculated from $\xi$ while circular symbols are from theoretical predictions.

The $\xi$ values obtained in ES-VRH allow us to estimate the bandgap ($E_g$) of RGO for different $sp^2$ fraction. From Kane model, relativistic spectrum $E(k)$ can be expressed as $E(k)^2 = E_g^2 + (\hbar v_F k)^2$, where $v_F$ is the graphene Fermi velocity. In the middle of the gap $E(k) = 0$, and the equation become $E_g^2 + (\hbar v_F k)^2 = 0$. For an imaginary $k = i/k/$, we can get $\xi = 1/|k| = \hbar v_F/E_g$.[61] Using the value of $\xi$ from Fig. 5(b), we calculated the values of $E_g$ as 1.43, 1.22, 0.64, 0.43, 0.29 and 0.21 eV for samples A, B, C, D, E and F respectively. These $E_g$ values are plotted against there corresponding carbon $sp^2$ fraction in Fig. 5 (c) (square symbols). We have also compared our $E_g$ values with that of theoretical $E_g$ (circular symbol) predicted by the DOS calculation.[62] The agreement between our experimental results and theoretical prediction are quite remarkable providing further evidence of the applicability of ES VRH for all of our RGO sheets.

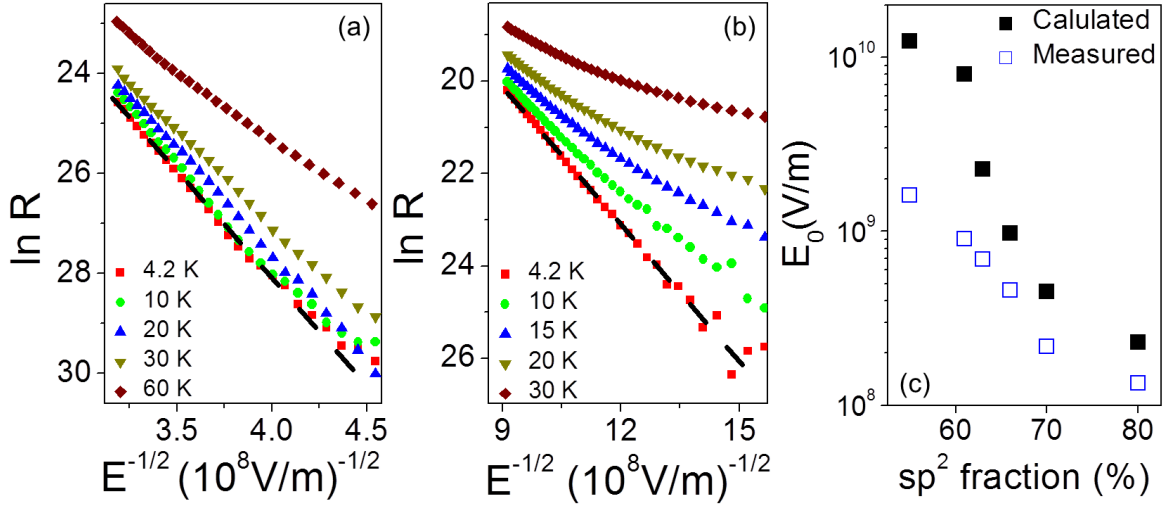

FIG. 6. (Color online) (a) $\ln R$ vs $E^{-1/2}$ for device A at temperature ranges from 4.2 to 60 K. (b) $\ln R$ vs $E^{-1/2}$ for device F at temperature ranges from 4.2 to 30 K. Dashed lines show a linear fit $E^{-1/2}$. (c) Comparison of hopping parameter $E_0$ determined from Ohmic and non-Ohmic ES-VRH with different carbon $sp^2$ fraction of RGO sheets. Solid symbols are calculated from Ohmic (low electric field regime) ES-VRH. Open symbols are found from experimental non-Ohmic (high electric field regime) ES-VRH.

Additional evidence of ES-VRH can be obtained from the high bias nonlinear $I$-$V$ curve. At high enough electric field (high bias regime) and low temperature, the temperature dependence is strongly reduced and one enters the regime of field driven hopping transport, where the conduction is given by Eq. (4). Figures 6(a) and (b) show $\ln R$ vs $E^{-1/2}$ characteristics of two representative devices A and F at a few selected temperatures down to 4.2 K. The value of $R$ is calculated by dividing the current with voltage. At higher temperatures, the curves are still temperature dependent within the measured bias voltage range (up to 5 V). However, as the temperature gets close to 4.2 K, the curves become weakly temperature dependent. It is possible to see temperature independent regime at higher bias voltage. However, we did not apply more than 5 V as the devices undergo electrical breakdown slightly above this voltage. We fitted the 4.2 K data with $E^{-1/2}$ (solid line) in high bias regime and the data fit very well, indicating that the $R$ follows field driven (or non-Ohmic) ES-VRH. Similar fits were also obtained for samples B, C, D and E (not shown here). It has been noted that the field dependent hopping equation is only valid when the electric field is higher than a critical field $E_C = 2k_B T / e\xi$.[39-42, 63] In our case this condition is satisfied as the values for $E_C$ at 4.2 K for device A and F are $16.7 \times 10^5$ V/m ($E_C^{-1/2}$ =7.74 $(10^8 \text{V/m})^{-1/2}$) and $2.28 \times 10^5$ V/m ($E_C^{-1/2}$ =20.9 $(10^8 \text{V/m})^{-1/2}$) and the fit was for $E^{-1/2} < E_C^{-1/2}$.

TABLE I. Summary of ES-VRH fitting results with varying carbon $sp^2$ fraction in RGO sheets.

| Devices (carbon $sp^2$ fraction (%)) | $R_{Room}$ (MΩ) | $R_0$ (kΩ) | $T_{ES}$ (K) | $\xi$ (nm) | $E_g$ (eV) | Calculated $E_0$ ($10^8$ V/m) | Measured $E_0$ ($10^8$ V/m) |
|---|---|---|---|---|---|---|---|
| A (55) | 1060 | 14.8 | 30976 | 0.46 | 1.43 | 123 | 16.1 |
| B (61) | 83.5 | 13.6 | 24964 | 0.54 | 1.22 | 80.1 | 9.05 |
| C (63) | 13.8 | 14.1 | 13280 | 1.03 | 0.64 | 22.7 | 6.91 |
| D (66) | 3.4 | 13.8 | 8704 | 1.54 | 0.43 | 9.73 | 4.60 |
| E (70) | 1.2 | 13.1 | 5901 | 2.27 | 0.29 | 4.47 | 2.19 |
| F (80) | 0.6 | 12.6 | 4225 | 3.21 | 0.21 | 2.29 | 1.38 |

From the slope of the fitted line in Figs. 6(a) and (b), we obtained the value of $E_0$ as 16.1 × $10^8$ and 1.38 × $10^8$ V/m for device A and F, respectively. The value of $E_0$ can also be calculated from Eq. (5) using the values of $T_0$ and $\xi$ obtained from the Ohmic ES-VRH. The corresponding values for $E_0$ were 123 × $10^8$ V/m and 2.29 × $10^8$ V/m for device A and F, respectively. Similar analysis was done for all other samples (B, C, D, and E) and the corresponding values of $E_0$ from Ohmic ES-VRH (marked as solid symbols) and experimentally measured values obtained from the slope in the high electric filed regime (marked as open symbols) are plotted against their corresponding carbon $sp^2$ fractions in Fig. 6(c). The results from two different regimes are in fairly good qualitative agreement. The small variation, also seen for ES-VRH in other materials, may indicate that the constants in Eqs. (3) and (5) may not be very accurate. A summary of all the results obtained from our measurements is presented in Table I.

## IV. CONCLUSION

We demonstrated ES-VRH in RGO sheets of varying carbon $sp^2$ fractions, both in Ohmic and non-Ohmic regime. In Ohmic regime, the temperature dependence of resistance for all the samples follow $R = R_0 \exp[(T_{ES}/T)^{1/2}]$ with $T_{ES}$ decreasing from 30976 to 4225 K and localization length $\xi$ increasing from 0.46 to 3.21 nm with increasing carbon $sp^2$ fraction from 55 to 80%. From the localization length, we calculate a bandgap variation of our RGO from 1.43 to 0.21 eV with increasing $sp^2$ fraction from 55 to 80 % which agrees remarkably well with theoretical prediction. At low temperature and high electric field (high bias regime), our data can be explained with field dependent ES-VRH model $R \sim \exp[(E_0/E)^{1/2}]$ with the values of $E_0$ obtained from the slope is in good agreement with that of $E_0$ obtained from the Ohmic regime. By extrapolating our data to 100 % $sp^2$ fraction, we conclude that Mott-VRH may not be observed in the chemically reduced RGO sheets.

## ACKNOWLEDGMENTS
We thank B.I. Shklovskii for many useful discussions and V. Singh for help with XPS analysis. This work has been partially supported by U.S. NSF under Grant No. ECCS 0748091 to S.I.K.